\documentclass[prx,twocolumn,showpacs]{revtex4}
\usepackage{amsthm}
\usepackage{amsmath}
\usepackage{latexsym}
\usepackage{amsfonts}
\usepackage{amssymb}
\usepackage{color}
\usepackage{bbm,dsfont}
\usepackage{graphicx}
\usepackage{hyperref}
\usepackage{subfigure}

\newcommand{\id}{\mathbb{I}}

\newcommand{\FCS}[1]{\rho^{\rm FC}_{#1}}
\newcommand{\cFCS}[1]{\rho^{\rm cFC}_{#1}}

\begin{document}

\title{Equivalence of matrix product ensembles of trajectories in open quantum systems}

\author{Jukka Kiukas}
\author{M\u{a}d\u{a}lin Gu\c{t}\u{a}}
\affiliation{School of Mathematical Sciences, University of
Nottingham, Nottingham, NG7 2RD, UK}
\author{Igor Lesanovsky}
\author{Juan P. Garrahan}
\affiliation{School of Physics and Astronomy, University of
Nottingham, Nottingham, NG7 2RD, UK}

\pacs{05.30.Rt,05.30.-d,64.70.P-}

\date{\today}

\begin{abstract}
The equivalence of thermodynamic ensembles is at the heart of 
statistical mechanics and central to our understanding of equilibrium 
states of matter. Recently, it has been shown that there is a formal 
connection between the dynamics of open quantum systems and the 
statistical mechanics in an extra dimension. This is established 
through the fact that an open system dynamics generates a Matrix Product 
state (MPS) which encodes the set of all possible quantum jump 
trajectories and permits the construction of generating functions in the 
spirit of thermodynamic partition functions. In the case of 
continuous-time Markovian evolution, such as that generated by a 
Lindblad master equation, the corresponding MPS is a so-called {\em 
continuous} MPS which encodes the set of continuous measurement records 
terminated at some fixed total observation time.  Here we show that if one instead 
terminates trajectories after a fixed total number of quantum jumps, 
e.g. emission events into the environment, the associated MPS is {\em 
discrete}. This establishes an interesting analogy: The continuous and 
discrete MPS correspond to different ensembles of quantum trajectories, 
one characterised by total time the other by total number of quantum 
jumps. Hence they give rise to quantum versions of different thermodynamic ensembles, 
akin to ``grand-canonical'' and ``isobaric", but for trajectories.
Here we prove that these trajectory ensembles are equivalent in a suitable 
limit of long time or large number of jumps. This is in direct analogy to equilibrium statistical 
mechanics where equivalence between ensembles is only strictly 
established in the thermodynamic limit.  An intrinsic quantum feature is that the equivalence holds only for all observables that commute with the number of quantum jumps. 
\end{abstract}

\maketitle

\section{Introduction}

There is currently much interest in achieving a fundamental understanding   of the non-equilibrium dynamics of quantum systems in general \cite{Polkovnikov2011,Nandkishore2014,Eisert2015}, and of open quantum systems in particular \cite{Plenio1998,Petruccione2002,Diehl2008,Rivas2012}.   When the dynamics of an open quantum system is, to a good approximation, Markovian (i.e.\ memoryless), its evolution is given by a master equation of the Lindblad type \cite{Lindblad1976}.  Such dynamics can be ``unravelled'' in terms of stochastic quantum trajectories \cite{Belavkin,Gardiner2004,Carmichael2009}.  Each quantum trajectory represents a different observation record (as measurements on the environment and consequently on the system, via the system-environment interaction).  Averaging over the whole ``ensemble'' of dynamical trajectories at any point in time gives the average system evolution described by the master equation \cite{Gardiner2004,Carmichael2009}.

The concept of trajectory ensembles is important as it allows to connect  \cite{Garrahan2010,Lesanovsky2013} open quantum dynamics to ideas of ensemble statistical mechanics \cite{Chandler1987,Peliti2011}.  Consider dynamics between zero initial time, up to some time $t$.  The dynamics generated by a Lindblad master operator is that of a continuous quantum Markov chain, one of quantum jumps occurring stochastically with dissipative, yet deterministic, evolution between the jumps.  The ensemble of trajectories is the set of all possible trajectories generated by this dynamics, each with an associated probability (or amplitude) of occurring and classified by one (or more) parameters that are held fixed, in this case the total time extent $t$ of the trajectories.  In a thermodynamic analogy (and for simplicity we use the language of classical systems) 
if we think of a (canonical) ensemble in a system of fixed volume at say infinite temperature, the ensemble is the set of all possible configurations of the system compatible with that volume, all occurring with equal probability (due to the infinite temperature).  

In thermodynamics a central role is played by order parameters and their conjugate fields \cite{Chandler1987,Peliti2011}.  Order parameters are system extensive observables that quantify properties of the system at the given conditions (e.g. density, magnetisation, etc.) which couple to their corresponding intensive fields (chemical potential, magnetic field, etc.).  In the case of quantum trajectories, order parameters are time-integrated quantities \cite{Garrahan2010,Lesanovsky2013}, the number of quantum jumps $K$ over a whole trajectory being one typical example.  
Since the trajectories are stochastic, a quantity such as $K$ will be random, and its distribution will encode the statistical properties of the dynamics.  In particular, interesting properties may be uncovered by considering atypical values of the order parameter, since often significant fluctuation behaviour is encoded in the tails of their distributions \cite{Garrahan2010,Lesanovsky2013}.  

This leads then to the idea of ``biased'' or ``tilted'' ensembles, i.e.\ ensembles conditioned on certain values of the order parameter \cite{Garrahan2010,Lesanovsky2013,Touchette2009}.  In the standard thermodynamic case, if for example the order parameter is the energy $E$, the ensemble of configurations with some fixed $E$ all with the same probability is termed microcanonical, while the ensemble of all possible configurations with probability proportional to $e^{-E/T}$, where $T$ is a fixed (scaled) temperature, is termed canonical.  The same can be done in the dynamical case by defining trajectory ensembles conditioned on, say, a certain fixed number of emissions $K$ (analogous to the microcanonical ensemble), or where $K$ is not fixed but trajectories are biased by a factor $e^{-s K}$ where $s$ is  a so-called counting field.  This ``thermodynamics of trajectory'' approach \cite{Garrahan2010,Lesanovsky2013}, sometimes called $s$-ensemble method \cite{Garrahan2007,Lecomte2007,Hedges2009}, allows to uncover key properties of the dynamics---often by means of large-deviation techniques \cite{Touchette2009} that become applicable when time is large---for example by revealing the existence of dynamical phase transitions  \cite{Garrahan2010,Lesanovsky2013,Garrahan2007,Lecomte2007,Hedges2009}.

A fundamental property of thermodynamic ensembles is that of {\em  equivalence}: for a given system the microcanonical and canonical ensemble are equivalent in the limit of large system size, as long as the temperature of the latter is such that the canonical average of the energy coincides with the microcanonical fixed value.  The same occurs with other ensembles, such as between those of fixed volume and fixed pressure.  The simplest manifestation of this equivalence is in the Legendre transform that connects the corresponding free-energies, i.e., a correspondence at the large-deviation level.  For dynamical trajectory ensembles, it has been shown \cite{Chetrite2013,Chetrite2014} that for classical stochastic dynamics the microcanonical (in the sense of fixed time-integrated observable) and canonical (fixed conjugate counting field) ensembles are indeed equivalent, as in the thermodynamics case.  Furthermore, in Ref.\ \cite{BTG14}, a new ensemble was introduced, one of trajectories have a fixed value of the time-integrated observable, such as the number of quantum jumps, but time is allowed to fluctuate (termed $x$-ensemble).  This is like the dynamical version of a fixed pressure (and thus fluctuating volume) ensemble, with $x$ playing the role of a pressure (on time).  In \cite{BTG14} the correspondence between the $s$-ensemble and $x$-ensemble was proved at the level of large-deviation rate functions, both for the classical and open quantum cases. 

In this paper we unify and extend these ideas about thermodynamics of trajectories and dynamical ensemble equivalence by using the technology of matrix product states (MPS).  We exploit the fact that the set of quantum trajectories of an open quantum system can be encoded in an MPS, and show how to do this for ensembles of trajectories with fixed number of jumps $K$.  By exploiting the MPS construct we prove in general the equivalence of microcanonical and canonical dynamical ensembles, thus generalising the classical results of \cite{Chetrite2013,Chetrite2014,CETT05,LPS94} to the quantum realm.  We also prove an equivalence result between the $s$-ensemble (fixed time) and the $x$-ensemble (fixed number of quantum jumps); in this case we consider the overlap between the pure (system together with output) states of the two corresponding canonical ensembles, and show that this overlap has subexpoential decay in the limit of large time. However, since in the time ensemble, the system has a nontrivial evolution after the last jump, we need to condition onto trajectories where the last jump happens at the end of the time interval. In an alternative approach, we compare the (mixed) states of the outputs of the two canonical ensembles, rather than the full system and output states; we find that the reduced states which describe the output up to a fixed time 
$\tau_0$ converge in the limit of large time and respectively counts, and provide explicit expressions of the limits. The equivalence is manifest here in that the $x$-ensemble limit is \emph{equal} to the diagonal part of the $s$-ensemble limit, with respect to the decomposition of the output Fock space into ``layers" of fixed number of counts subspaces. However, the reduced state of the $x$-ensemble does not have any coherence between different layers, which may be present however in the $s$-ensemble. We illustrate this purely quantum feature by considering the dependence of the two outputs on specific dynamical parameters; 
in particular the $s$-ensemble is shown to be sensitive to phase transformations generated by the counting operator, while the $x$-ensemble is left invariant due to the above mentioned lack of coherence.


The paper is organised as follows.  In Sect.\ II we set up the formalism describing open dynamics in terms of MPS.  In order to maintain as much generality as possible we discuss open systems initially in terms of continual quantum measurements.  This allows to capture both the standard situation of open dynamics observed for a fixed total amount of time, and also the case where observations are for a fixed number to quantum jump.  While the former is known to be encoded in so-called continuous MPS for the system and output, we show that the later is encoded discrete MPS states, and describe the connection between the two.  In Sect.\ III we describe the thermodynamics of trajectories approach, and discuss both the $s$-ensemble and $x$-ensemble methods in the context of MPS for the system-output.  Section IV contains our key results.  There we prove the equivalence of ensembles.  We provide the proof in three stages.  The first one is the correspondence of potentials, or large-deviation rate functions, which duly follows from relations between the tilted superoperators that define the ensembles.  The second stage involves proving the actual equivalence of measures over quantum trajectories, in analogy to what is done in the classical case.  The final stage is the proof of full operational equivalence between the ensembles, on the level of diagonal blocks with respect to Fock layers.  In Sect.\ V we discuss the specific case of quantum renewal processes, where the fixed number of jumps ensemble becomes particularly simple and transparent, and illustrate it with an example.  
Finally in Sect.\ VI we provide our conclusions and outlook.

\section{Continual quantum measurements and Matrix Product States}

We consider the scenario where a quantum system with Hilbert space $\mathcal{H}$ is continuously monitored by performing measurements on it (typically realised via coupling the system with an external environment.) In the counting case which we describe below, the measurement record  consists of clicks of different types, occurring at random times, according to a certain probability distribution  \cite{Da76} (see also \cite{We87,KW12} for a viewpoint closer to ours).

Suppose that the system starts in an initial state $\psi\in \mathcal{H}$. For a random \emph{waiting time} $t_w$ the system evolves continuously, so that its unnormalised state at time $t<t_w$ is $\exp(-iH_{\rm eff} t) \psi$, where $H_{\rm eff}$ is a non-selfadjoint \emph{effective Hamiltonian}. The probability density of the waiting time is given by the loss of wavefunction normalisation 
$p_\psi(t_w):=-\frac{d}{dt}\|e^{-it H_{\rm eff}}\psi\|^2|_{t=t_w}$.
 At the time of detection, a click with label $i=1,\ldots,N_L$ is recorded, and the system's conditional state is updated by applying a jump operator $L_i$. The full measurement is therefore described by the positive operator valued measure (POVM) $(t_w,i)\mapsto J[t_w,i]^\dagger J[t_w,i]$, where $J[t_w,i]=L_ie^{-it_wH_{\rm eff}}$ is the \emph{jump operator} effecting the total change $\psi\mapsto J[t_w,i]\psi$. These must satisfy
\begin{align}\label{normalisation}
p_\psi(t_w)&=\sum_{i=1}^{N_L}\|J[t_w,i]\psi\|^2, & \int_0^\infty dt_w\, p_\psi(t_w)=1,
\end{align}
the first condition stating that a detection event happening at waiting time $t_w$ necessarily involves one of the outcomes $i=1,\ldots,N_L$ being realised, and the second fixing the total normalisation of waiting time. From these conditions it follows that $H_{\rm eff}=H-\frac{i}{2}\sum_{i}L_i^*L_i$, where $H$ is a selfadjoint operator interpreted as the system's Hamiltonian when isolated from the environment.

After the first detection, the process is repeated starting with the state $J[t_w,i]\psi / \| J[t_w,i]\psi\|$, with an independent waiting time. A full detection process is given by a finite measurement \emph{trajectory} ${\bf X}=((t_1,i_i),\ldots, (t_n,i_n))$, where $0\leq t_1\leq\cdots\leq t_n$. Each such trajectory has the \emph{final time} $T[{\bf X}]=t_n$, and \emph{total number of jumps} $K[{\bf X}]=n$. When integrating over trajectories we will use the notation
\begin{equation}\label{integral}
\int d{\bf X} =\sum_{n=0}^\infty\sum_{i_1,\ldots,i_n}\int_{0\leq t_1\leq\cdots\leq t_n} dt_1\cdots dt_n.
\end{equation}
Since we are dealing with quantum systems, it will turn out to be convenient to understand classical trajectories ${\bf X}$ as labels for quantum states $|{\bf X}\rangle$ spanning an abstract \emph{output space} $\mathcal F^{\rm out}$ of wavefunctions $$\Psi=\int d{\bf X}\, \langle {\bf X}|\Psi\rangle\, |{\bf X}\rangle,$$ with orthogonality relations $\langle {\bf X}|{\bf X}'\rangle=\delta({\bf X}-{\bf X}')$
matching with the integral \eqref{integral}. Due to the ordering of the time points, the wavefunctions $\Psi({\bf X})=\langle {\bf X}|\Psi\rangle$ correspond uniquely to symmetric functions of $K[{\bf X}]$ unconstrained time variables on $[0,\infty)$. Hence, we may identify the output space with the Bosonic Fock space 
\begin{equation}\label{eq.Fock.space}
\mathcal F^{\rm out} \cong 
\bigoplus_{n=0}^\infty
L^2\left([0,\infty),\mathbb{C}^{N_L}\right)^{\otimes_s n} 
\end{equation}
If $a_i(t)$ are the (singular) Bosonic annihilation operators at time $t$, and $|\Omega\rangle$ is the vacuum state, then $|{\bf X}\rangle=\hat a_{i_1}^\dagger(t_1)\cdots \hat a^\dagger_{i_{K[{\bf X}]}}(T[{\bf X}])|\Omega\rangle/{\rm norm}$. This conveys the intuitive idea that each detection event corresponds to a ``particle" in the output.  

From the experimental point of view, there are two natural ways of obtaining finite trajectories from the above scheme; the distinction has not been made explicit in the existing literature, and our first aim in the present paper is to clarify it. The first one repeats the single detection process $K$ times. Clearly this is particularly relevant when the detectors have to be initialised after each detection, so that each measurement is \emph{actively performed} on the system.  This scheme has an associated state transformation given by,
\begin{align*}
J[{\bf X}] &=J(t_{w,n}, i_n)\cdots J(t_{w,2},i_2)J(t_{w,1},i_1) ,
\end{align*}
where $t_{w,n}=t_n-t_{n-1}$, $t_{w,1}=t_1$ are the waiting times between 
the detection events, or quantum jumps.  In words, given a trajectory ${\bf X}$ resulting from this process, the system is at the end in state $J[{\bf X}]\psi$. 

The second scheme for trajectories is to let the process run until a given final time $\tau$. This suits better situations where the system is evolving in time, and detections take place spontaneously. The associated state transformation in this case is,
\begin{align}\label{v.tau}
V^\tau[{\bf X}] &=e^{-i(\tau-T[{\bf X}]) H_{\rm eff}}J[{\bf X}] .
\end{align}
In this case, given a trajectory ${\bf X}$ resulting from the process terminated at time $\tau$, puts the system in the final state $V^\tau[{\bf X}]\psi$ due to the contractive evolution after the final jump. 

We now construct suitable output states describing these two processes, and relate them to the evolution of open quantum systems.

\subsection{Measurements with fixed number of jumps}

We fix a number of jumps $K$, and repeat the single detection process $K$ times. For each realisation, we record the trajectory ${\bf X}$, together with the corresponding conditional system evolution $J[{\bf X}]\psi$, and arrange them into a quantum superposition over all possible realisations:
\begin{equation}\label{MPS}
|{\rm MPS}_K\rangle=\int_{{\bf X}:K[{\bf X}]=K} d{\bf X}\, J[{\bf X}]|\psi\rangle\otimes |{\bf X}\rangle.
\end{equation}
This pure state belongs to subspace of $\mathcal{H}\otimes \mathcal{F}^{\rm out}$ consisting of the tensor product between system and the $K$-th layer of the output space \eqref{eq.Fock.space}, and it is easy to check that it is normalised due to \eqref{normalisation}. Its reduced density matrix \emph{on the output alone} will be denoted by $\FCS{K}$, which is finitely correlated \cite{FNW92}. Its matrix elements are $\langle {\bf X}'|\FCS{K}|{\bf X}\rangle ={\rm tr}[\rho J[{\bf X}]^\dagger J[{\bf X}']]$, where $\rho= |\psi\rangle\langle\psi |$. 
For ${\bf X}={\bf X}'$, this is just the probability density for the trajectory ${\bf X}$.

If we decouple the system at the end via suitable measurement (instead of tracing over it), we get a pure state, which is of matrix product form \cite{PVW07} with bond dimension at most the dimension of the system. This can be seen as follows: we first map the $K$-layer of the output space unitarily to the $K$-fold tensor product of the Hilbert space with waiting time basis $|t_{w},i\rangle$, where $0\leq t_w\leq \infty$, $i=1,\ldots N_L$, by making a (unitary) identification
$$|(t_1,i_1),\ldots,(t_K,i_K)\rangle\simeq |t_{w,1},i_1\rangle\otimes\cdots\otimes |t_{w,K},i_K\rangle.$$
We then choose a basis $\{|m\rangle\}$ for the \emph{system}, and define normalizable \emph{output} vectors
$$
\Psi_{m,m'}=\int_0^\infty dt_w \sum_{i=1}^{N_L} \langle m|J(t_w,i)|m'\rangle |t_w,i\rangle.
$$
Now clearly
\begin{align}
|{\rm MPS}_K\rangle &= \sum_{m_0,\ldots,m_K} |m_K\rangle\langle m_0|\psi\rangle\otimes\nonumber\\
&\Psi_{m_K,m_{K-1}}\otimes \cdots \otimes\Psi_{m_2,m_1} \otimes \Psi_{m_1,m_0},\label{waitingrep}
\end{align}
which can easily be put into an MPS form by choosing an orthonormal basis for the linear span of the vectors $\Psi_{m,m'}$. Note that this just amounts to diagonalising the associated Gram matrix
$$
G_{m,m';\tilde m,\tilde m'} = \int_0^\infty dt_w \sum_{i=1}^{N_L} \langle m'|J(t_w,i)^\dagger |m\rangle \langle \tilde m|J(t_w,i) |\tilde m'\rangle.
$$
We call \eqref{waitingrep} the waiting time representation of the state $|{\rm MPS}_K\rangle$.

It is also useful to find the \emph{unconditional evolution} of the system under the measurement process, obtained by averaging over all realisations. This is simply obtained from $|{\rm MPS}_K\rangle$ by tracing over the output:
\begin{align*}
{\rm tr}_{\rm out}[|{\rm MPS}_K\rangle\langle {\rm MPS}_K|] &=\int_{{\bf X}:K[{\bf X}]=K} d{\bf X}{\rm tr}[J[{\bf X}]\rho J[{\bf X}]^\dagger]\\
&= \mathbb T_*^K(\rho),
\end{align*}
where $\mathbb T_*$ is the Schr\"odinger picture version of the quantum channel
$$\mathbb T(\cdot)=\sum_{i=1}^{N_L}\int_0^\infty dt_w J[t_w,t]^\dagger(\cdot)J[t_w,i],$$
describing one measurement step in the Heisenberg picture. Since the process is discrete, the subsequent evolution is obtained by iterating $\mathbb T$.
\subsection{Measurements with fixed final time}

When the observation is terminated at fixed time $\tau$, each realisation consists of a trajectory ${\bf X}$ with $T[{\bf X}]\leq \tau$, and the corresponding operator $V^\tau[{\bf X}]$. We record this information into the state
\begin{equation}\label{cMPS}
|{\rm cMPS}_\tau\rangle:=\int_{{\bf X}:T[{\bf X}]\leq \tau} d{\bf X}\, V^\tau[{\bf X}]\psi \otimes |{\bf X}\rangle.
\end{equation}
We recall that the output Fock space \eqref{eq.Fock.space} satisfies the factorisation property 
$\mathcal F^{\rm out} \cong \mathcal F^{\rm out}_{(0,\tau]}\otimes \mathcal F^{\rm out}_{(\tau,\infty)}$ where each term is a Fock space of trajectories up to and respectively after time $\tau$. Under this factorisation, the above state belongs to the subspace  
$\mathcal{H}\otimes \mathcal{F}^{\rm out}_{(0,\tau]} \otimes \Omega_{(\tau,\infty)} \subset \mathcal{H}\otimes \mathcal{F}^{\rm out}$ of system and output states with ``spatial volume'' restricted by the final time $\tau$. 
We note that the state can be written (see e.g. \cite{HCOV10}) implicitly in terms of a path ordered exponential:
$$
|{\rm cMPS}_\tau\rangle=\mathcal Pe^{\int_0^\tau dt \left(H_{\rm eff}\otimes \id_{\rm field}+\sum_i L_i\otimes \hat{a}_i^\dagger(t)\right)}(|\psi\rangle\otimes |\Omega\rangle).
$$

Conditional states obtained from this state by decoupling the system via suitable measurements, are called Continuous MPS (cMPS), which were relatively recently introduced as useful variational states for quantum field theory \cite{VC10, OEV10}. For later purposes we denote the output reduced state by $\cFCS{\tau}$.

The averaged system evolution is now 
\begin{align*}
{\rm tr}_{\rm out}[|{\rm cMPS}_\tau\rangle\langle {\rm cMPS}_\tau|]&=\int_{{\bf X}:T[{\bf X}]\leq \tau} d{\bf X} {\rm tr}[V^\tau[{\bf X}]\rho V^\tau[{\bf X}]^\dagger]\\
&= e^{\tau \mathbb W_*}(\rho),
\end{align*}
where the last equality is the Dyson expansion of the exponential $e^{\tau \mathbb W_*}$ with
$$
\mathbb W(\cdot)=-\mathcal R(\cdot)+\sum_{i=1}^{N_L} L_i^\dagger (\cdot)L_i,
$$
and $\mathcal R(\cdot)=i(\cdot)H_{\rm eff}-iH_{\rm eff}^\dagger(\cdot)$. Of course, this is the Lindblad generator of a continuous Markovian open system evolution (semigroup of channels) $\tau\mapsto e^{\tau \mathbb W}$. In physics literature, the measurement process is often called (quantum jump) unravelling of the evolution \cite{MC96}. We have here presented it ``the other way around" so as to emphasise its similarities and differences to the above fixed $K$ scenario, which develops more naturally from the jump picture, and has not been extensively investigated in the literature so far.

From the mathematical point of view, the construction of the abstract output space is an instance of the measurement dilation theory of Naimark (cf. e.g. \cite{Pa03}) and Davies \cite{Da76}, which states that counting processes arising from generalised (POVM) measurements can always be realised projectively on a larger Hilbert space. In concrete examples, the Lindblad generator often results from a unitary evolution coupling the system with some physical environment, typically a Bosonic heat bath. In this case, the abstract output field $\mathcal F^{\rm out}$ can naturally be identified with the environment, and the measurements themselves can be realised physically by measuring suitable field operators. This leads to input-output models for optical systems and stochastic Schr\"{o}dinger equations \cite{Gardiner2004} or filtering equations \cite{Belavkin} describing the conditional evolution of the system monitored through the environment.

\begin{figure*}[t!]
\includegraphics[width=2\columnwidth]{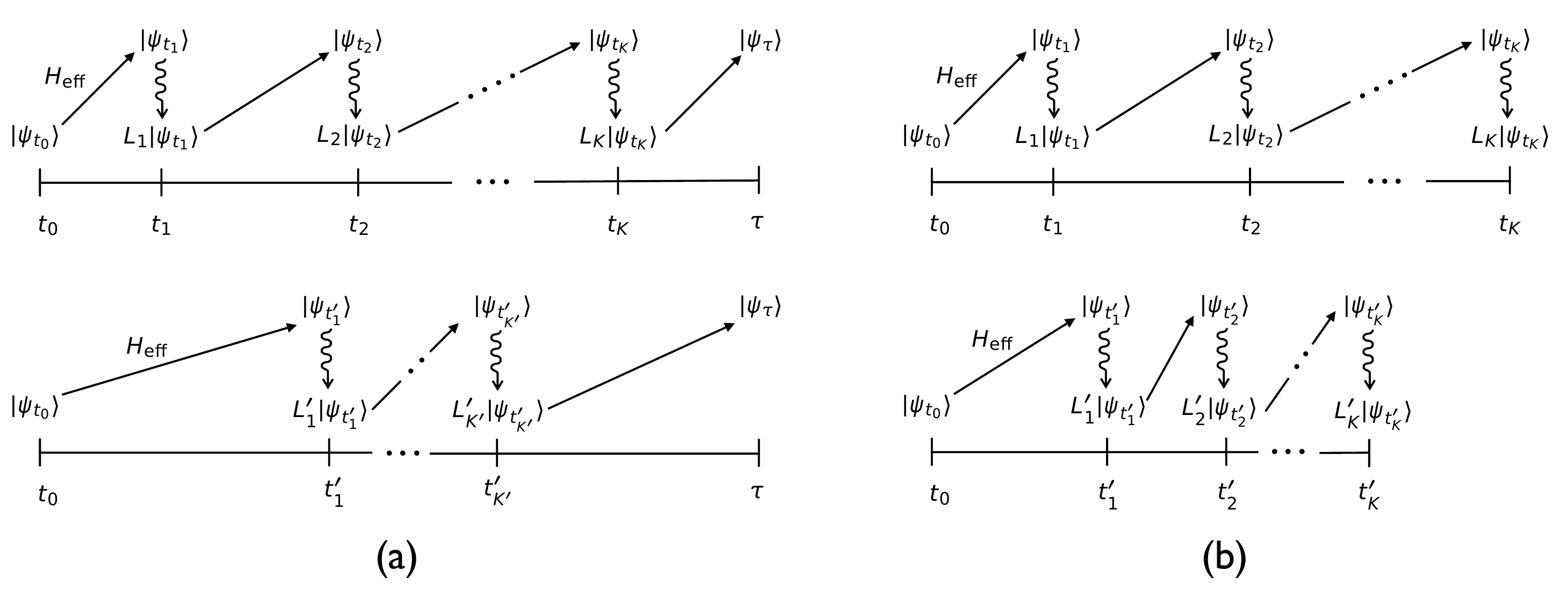}
\caption{Sketch of trajectory ensembles.  Time runs from left to right.  
Wiggly lines indicate quantum jumps, and straight ones deterministic evolution between jumps under $H_{\rm eff}$.  (a) The $s$-ensemble.  All trajectories in the ensemble are of a total fixed time $\tau$ but can have any number of quantum jumps.  Furthermore, between the time of the last jump and the final time there may be a period of no-jump evolution.  (b) The $x$-ensemble.  All trajectories in the ensemble have the same number of quantum jumps $K$, but their time extension can fluctuate.  In this case trajectories terminate after the $K$-th jump.}
\label{fig1}
\end{figure*}

\section{Thermodynamics of trajectories}\label{secIII}

Viewing the abstract output as a physical field with trajectories corresponding to configurations of excitations or ``particles" on the output allows us to employ techniques from non-equilibrium statistical mechanics (see e.g. \cite{Ru04}) to study their properties at the limit of long observation times. The mathematical underpinning is the large deviations theory for the relevant extensive observables, which we now proceed to describe.

We are interested in quantities obtained by incrementing with some amount at the addition of each particle. Such a quantity is of the form $F[{\bf X}]=\sum_{n=1}^{K[{\bf X}]} F(t_{w,n},i_n)$, where $F(t_w,i)$ is a (possibly vector valued) quantity depending only on a single waiting time $t_w$, and some property of the system we are monitoring, say ``spin" $i$. Such quantity can be represented as the diagonal field operator
\begin{align*}
\hat F &=\int d^K{\bf X}\,\, F[{\bf X}]\, |{\bf X}\rangle\langle {\bf X}|,
\end{align*}
and the expectation values of any function $f(F)$ in the above two processes have quantum expressions ${\rm tr}[\FCS{K} f(\hat F)]$ and ${\rm tr}[\cFCS{\tau} f(\hat F)]$. In particular, by taking $f(F)=\delta(F-F_0)$ gives the probability distributions of the observable $\hat F$ at point $F_0$.

The most important time-extensive quantities are the particle number $\hat K$ and total volume $\hat T$, obtained by taking $F(t_w, i)=1$ and $F(t_w,i)=t_w$, respectively. These correspond to the above two ways of truncating an infinite trajectory, leading to statistical ensembles of field particles with either fixed particle number $K$ and fluctuating $T$, or fixed volume $T$ and fluctuating $K$. In both cases we can also measure a ``spin" operator $\hat M$ corresponding to $F(t_w,i)=M(i)$, where $M$ is some (in general vector valued) quantity depending on $i$.
The associated probability distributions are given by
\begin{align*}
P_K(T,M)&:={\rm tr}\left[\FCS{K}\delta\left((T,M)-(\hat T,\hat M)\right)\right],\\
P_\tau(K,M) &:={\rm tr}\left[\cFCS{\tau}\delta\left((K,M)-(\hat K,\hat M)\right)\right].
\end{align*}
Our subsequent analysis rests crucially on the properties of the associated generating functions
\begin{align*}
Z_K(x,c)&:={\rm tr}[\FCS{K} e^{-x\hat{T}-c\cdot \hat M}]={\rm tr}[\rho \mathbb T_{x,c}^K(\id)],\\
Z_\tau(s,c) &:={\rm tr}[\cFCS{\tau}e^{-s\hat{K}-c\cdot \hat M}]={\rm tr}[\rho e^{\tau \mathbb W_{s,c}}(\id)],
\end{align*}
written in terms of the \emph{deformed generators} \cite{BTG14} $\mathbb T_{x,c}$ and $\mathbb W_{s,c}$, obtained from $\mathbb T$ and $\mathbb W$ by replacing $J[{\bf X}]$ and $V^\tau[{\bf X}]$ with
\begin{align*}
J_{x,c}[{\bf X}] &=J_{x,c}(t_{w,n}, i_n)\cdots J_{x,c}(t_{w,2},i_2)J_{x,c}(t_{w,1},i_1),\\
V_{s,c}^\tau[{\bf X}] &=e^{-\frac 12 K[{\bf X}]s}e^{-i(\tau-T[{\bf X}]) H_{\rm eff}}J_{0,c}[{\bf X}],
\end{align*}
respectively, where $J_{x,c}(t_w,i)=e^{-\frac 12(xt_w+c\cdot M(i))}J(t_w,i)$. Explicitly, we have
\begin{align}
\mathbb T_{x,c}&=(x\,{\rm Id}+\mathcal R)^{-1}\left(\sum_{i=1}^{N_L} e^{-c\cdot M(i)}L_i^\dagger(\cdot)L_i\right),\label{Txc}\\
\mathbb W_{s,c}&=-\mathcal R(A)+e^{-s}\sum_{i=1}^{N_L} e^{-c\cdot M(i)}L_i^\dagger(\cdot)L_i,\label{Wsc}
\end{align}
where the inverse $$(x\,{\rm Id}+\mathcal R)^{-1}=\int_0^{\infty} dt (e^{-it(H_{\rm eff}-ix/2)})^\dagger(\cdot)e^{-it(H_{\rm eff}-ix/2)}$$ exists whenever
$\|e^{-it(H_{\rm eff}-ix/2)}\|\leq 1$,
which holds for all $x>x_{\rm min}$, where $x_{\rm min} =2{\rm Im}\lambda_0\leq 0$ and $\lambda_0$ is the eigenvalue of $H_{\rm eff}$ with maximum imaginary part \footnote{Note that $x_{\rm min}\leq 0$ because $\|e^{-itH_{\rm eff}}\|\leq 1$. In the generic case, $x_{\rm min}<0$, in which case the generating functions are analytic in a neighbourhood of the origin.}. We restrict to $x>x_{\rm min}$ subsequently.

We now suppose that $\mathbb T_{x,c}$ has a unique eigenvalue $e^{g(x,c)}$ equal to its spectral radius, and that $\mathbb W_{s,c}$ has a unique eigenvalue $\theta(s,c)$ with largest real part. We further assume that both eigenvalues are nondegenerate, associated to strictly positive eigenvectors $F_{x,c}$ and $E_{s,c}$, respectively, and that they are the \emph{only} eigenvalues with positive eigenvectors. We then assume that $e^{g(x,c)}$ and $\theta(s,c)$ are also the dominant eigenvalues of the corresponding Schrodinger picture generators $(\mathbb T_{x,c})_*$ and $(\mathbb W_{s,c})_*$ in the same sense, with unique eigenvectors $\sigma_{x,c}$ and $\sigma_{s,c}$. These conditions hold under the generic assumption of \emph{strong irreducibility (primitivity)} of the generators \footnote{Concerning the discrete case, the map $\mathbb T_{x,c}$ is irreducible if there is no projection $P$ with $\mathbb T_{x,c}(P)\leq \lambda P$ for some $\lambda$. Then the noncommutative Perron-Frobenius Theorem (see e.g. \cite[Thm. 2.4]{evans}) guarantees that the stated assumptions are fulfilled, except the uniqueness of the eigenvalue with maximum modulus $e^{g(x,c)}$. With this additional assumption, the map is called primitive.}, but this is not a necessary condition (see Sec. \ref{sec:renewal}).

With the usual normalisation (see e.g. \cite{JS10}) ${\rm tr}[\sigma_{x,c}]={\rm tr}[\sigma_{s,c}] ={\rm tr}[F_{x,c}\sigma_{x,c}]={\rm tr}[E_{s,c}\sigma_{s,c}]=1$, we then have
\begin{align}
e^{-Kg(x,c)} \mathbb T_{x,c}^K\xrightarrow{K\rightarrow\infty} {\rm tr}[\sigma_{x,c}(\cdot)]F_{x,c}\label{LDZK}\\
e^{-\tau \theta(s,c)} e^{\tau \mathbb W_{s,c}}\xrightarrow{\tau\rightarrow\infty} {\rm tr}[\sigma_{s,c}(\cdot)]E_{s,c}.\label{LDZT}
\end{align}
In particular, the generating functions satisfy $Z_K(x,c)\sim e^{K\, g(x,c)}$ and $Z_\tau(s,c) \sim e^{\tau \theta(s,c)}$ at these limits. The extensive parameters $K$ or $\tau$ being large indicates thermodynamic limit at the output field. The above mathematical result that $g(s,c)$ and $\theta(s,c)$ are well-defined in this limit lets us interpret them as \emph{thermodynamic potentials}, and $Z_K(x,c)$ and $Z_\tau(s,c)$ represent \emph{path canonical partition functions} \footnote{This terminology has been used in the classical case in \cite{Chetrite2013}.}, with the intensive variables $x$, $s$, and $c$ acting as \emph{control parameters} for the extensive quantities $T$, $K$, and $M$, respectively. In particular, we can identify $x$ with pressure (of the output excitations), and $s$ the chemical potential (controlling creation of new excitations).

We proceed with the thermodynamic analogy by interpreting the partition functions as normalisation factors for the corresponding \emph{path canonical ensembles}, given by quantum states
\begin{align}
&|{\rm MPS}_{K;x,c}^{\rm can}\rangle:=
Z_K(x,c)^{-\frac 12}e^{-\frac 12(x\hat T+c\cdot \hat M)}|{\rm MPS}_K\rangle\nonumber \\
&= \frac{1}{Z_K(x,c)^{\frac 12}}\int_{{\bf X}:K[{\bf X}]=K} d{\bf X}\, J_{x,c}[{\bf X}]|\psi\rangle\otimes |{\bf X}\rangle\label{canMPS}\\[2mm]
&|{\rm cMPS}^{\rm can}_{\tau;s,c}\rangle:=
Z_\tau(s,c)^{-\frac 12}e^{-\frac 12(s\hat K+c\cdot \hat M)}|{\rm cMPS}_\tau\rangle\nonumber\\
&=\frac{1}{Z_\tau(s,c)^{\frac 12}}\int_{{\bf X}:T[{\bf X}]\leq \tau} d{\bf X}\, V_{s,c}^\tau[{\bf X}]\psi \otimes |{\bf X}\rangle
\label{cancMPS}.
\end{align}
We denote the corresponding reduced output states by $\FCS{K;x,c}$ and $\cFCS{\tau;s,c}$; the corresponding classical trajectory ensembles $\langle {\bf X}|\FCS{K;x,c}|{\bf X}\rangle$ and $\langle {\bf X}|\cFCS{\tau;s,c}|{\bf X}\rangle$ are sometimes called \emph{$x$-ensemble} and \emph{$s$-ensemble}, respectively \cite{BTG14}.

From the asymptotic behaviour of the partition functions it follows by the Gartner-Ellis theorem that the corresponding probability distributions satisfy the LD principle, when properly scaled:
\begin{align}
P_K(T,M) & \sim e^{-K\, \phi(T/K,M/K)}, & K \text{ large } \label{LDprobK}\\
P_\tau(K,M) & \sim e^{-\tau \varphi(K/\tau,M/\tau)}, &\tau \text{ large}.\label{LDprobT}
\end{align}
These functions represent \emph{microcanonical partition functions} for the above thermodynamical ensembles on the output field, determined by fixed extensive variables $T,M,K$. The associated thermodynamic potentials $\phi(t,m)$ and $\varphi(k,\tilde m)$ only depend on the appropriately scaled intensive quantities $t=T/K$ (total volume per excitation) and $m=M/K$ (average value of $M$ per excitation) for the ensemble of fixed $K$, as well as $k=K/\tau$ (total number of excitations per unit volume) and $\tilde m=M/\tau$ (average value of $M$ per unit volume) for the ensemble of fixed volume. The associated \emph{path microcanonical states} are 
\begin{eqnarray}
&&
|{\rm MPS}^{\rm mc}_{K;T,M}\rangle 
:= 
\frac{\delta((\hat T,\hat M)-(T,M))|{\rm MPS}_K\rangle}{\sqrt{P_K(T,M)}} \nonumber\\
&&=\frac{1}{\sqrt{P_K(T,M)}} \int_{{\bf X} \in  \mathcal{T}_{K; T,M } }J[{\bf X} ] |\psi\rangle\otimes |{\bf X} \rangle 
\label{eq.microcanonical.K.tau}
\end{eqnarray}
where $ \mathcal{T}_{K;T,M}:= \{{\bf X} :  K[{\bf X}]=K , T[{\bf X}] = T , M [{\bf X}] =M \}$ 
is the space of trajectories with $K$ jumps, the last jump occurring at time $T$, and the "spin" observable taking value 
$M$, and
\begin{eqnarray}
&&
|{\rm cMPS}^{\rm mc}_{\tau; K,M}\rangle 
:= 
\frac{\delta((\hat K,\hat M)-(K,M))|{\rm cMPS}_\tau\rangle}{\sqrt{P_\tau(K,M)} } \nonumber
\\
&&= \frac{1}{\sqrt{P_\tau (K,M)}}
\int_{{\bf X} \in  \mathcal{T}_{\tau;K,M } }V^\tau[{\bf X}]\psi \otimes |{\bf X}\rangle. 
\label{eq.microcanonical.tau.K}
\end{eqnarray}
where $\mathcal{T}_{\tau; K,M} := \{ {\bf X} : T[{\bf X}] \leq \tau,  K[{\bf X}]=K , M [{\bf X}] =M\}$. 
Note that the microcanonical states are singular vectors, when seen as functions over \emph{all} trajectories, but can also be viewed as bona-fide Hilbert space functions over the restricted sets of trajectories, and are normalised in the latter sense.

The potentials $g$ and $\phi$ are connected to each other by the Legendre transform, and so are $\theta$ and $\varphi$. Assuming that the functions are smooth and convex, we can apply the usual machinery of classical thermodynamics to get
\begin{align}
g(x,c)&=-\phi(t,m)-tx-m\cdot c, \label{legendrex}\\
\theta(s,c) &=-\varphi(k,\tilde m)-ks-\tilde m\cdot c,\label{legendres}
\end{align}
where
\begin{align*}
t(x,c)&=-\frac{\partial g}{\partial x} =\lim_{K\rightarrow\infty}\frac{\langle T[{\bf X}]\rangle_{K,x}}{K},\\
m(x,c) &=-\frac{\partial g}{\partial c_j} =\lim_{K\rightarrow\infty}\frac{\langle M_j[{\bf X}]\rangle_{K,x}}{K},\\
k(s,c)&=-\frac{\partial\theta}{\partial s} =\lim_{\tau \rightarrow\infty}\frac{\langle K[{\bf X}]\rangle_{\tau,s}}{\tau},\\
\tilde m(s,c)&=-\frac{\partial\theta}{\partial c_j} =\lim_{\tau \rightarrow\infty}\frac{\langle M_j[{\bf X}]\rangle_{\tau,s}}{\tau},
\end{align*}
and the brackets denote expectation value with respect to the associated canonical ensembles.
These familiar relations describe the thermodynamics of the output field, in terms of the relevant intensive quantities obtained from the underlying continual measurement process. 
\section{Equivalence of ensembles}
In classical statistical mechanics, the concept of \emph{equivalence of ensembles} appears in different levels. Here we look at analogous statements for our quantum setup, the connection between the two different measurement scenarios, fixed $K$ and fixed $\tau$.

\subsection{Equivalence of potentials}
Often the equivalence is stated on the level of thermodynamic potentials. We have already seen that the Legendre transformations \eqref{legendrex} and \eqref{legendres} establish one-to-one correspondence between the potentials for path canonical and microcanonical partition functions, \emph{separately} for systems of fixed $K$ and fixed $\tau$. In \cite{BTG14}, it was shown that there is also a one-to-one-correspondence between potentials $g(x,c)$ and $\theta(s,c)$. Since that result will be essential also for our subsequent development, we present it here in more detail.

From \eqref{Txc} and \eqref{Wsc} it follows that
\begin{equation}\label{connection}
\mathbb T_{x,c}-e^s\,{\rm Id}=e^s(x\,{\rm Id}+\mathcal R)^{-1}\circ(\mathbb W_{s,c}-x\,{\rm Id}),
\end{equation}
holding for all admissible values of the parameters. This connection between the generators is the fundamental reason for the ensemble equivalence. It implies that for a fixed $c$ and $x$, the eigenvector $F_{x,c}$ of $\mathbb T_{x,c}$ corresponding to the dominant eigenvalue $e^{g(x,c)}$, is also an eigenvector of $\mathbb W_{s,c}$, with eigenvalue $x$, where $s=g(x,c)$. By the uniqueness of positive eigenvectors (see Sec. \ref{secIII}), it follows that
\begin{align}
x&=\theta(s,c), & s&=g(x,c),\label{connectionxs}\\
F_{x,c}&=\alpha_{s,c} E_{s,c}, && \text{ with some } \alpha_{s,c}>0.\label{connectionFE}
\end{align}
Using the above thermodynamic relations, together with \eqref{connectionxs}, we see that the intensive quantities defined via systems of fixed $K\rightarrow\infty$ are related to those defined via systems of fixed $\tau\rightarrow\infty$ by
\begin{align}
t(x,c)&=1/k(s,c), & m(x,c)&=\tilde m(s,c)/k(s,c).\label{tkconnection}
\end{align}
These relations show how the two different ensembles correspond to each other. In particular, the first relation in \eqref{tkconnection} states that $T/K$ (total volume per excitation), for a given pressure $x$ in the ensemble of fixed $K$, is equal to the inverse of $K/T$ (number of excitations per total volume), in the ensemble of fixed $T$, \emph{for a specific chemical potential $s$ uniquely determined by $x$}. 

By taking the adjoint of \eqref{connection}, we get the corresponding relation for the Schr\"odinger picture generators $(\mathbb T_{x,c})_*$ and $(\mathbb W_{s,c})_*$. The dominant eigenvalues are again $e^{g(x,c)}$ and $\theta(s,c)$, respectively (by our assumptions in Sec. \ref{secIII}), so the same correspondence holds. However, the corresponding eigenvectors $\rho_{x,c}$ and $\rho_{s,c}$ (normalised to having trace one) are \emph{not} equal, but related via
\begin{equation}\label{connectionev}
\rho_{s,c}=(x {\rm Id}+\mathcal R_*)^{-1}(\rho_{x,c})/ \text{ trace }.
\end{equation}

\begin{figure}[t!]
\includegraphics[width=0.9\columnwidth]{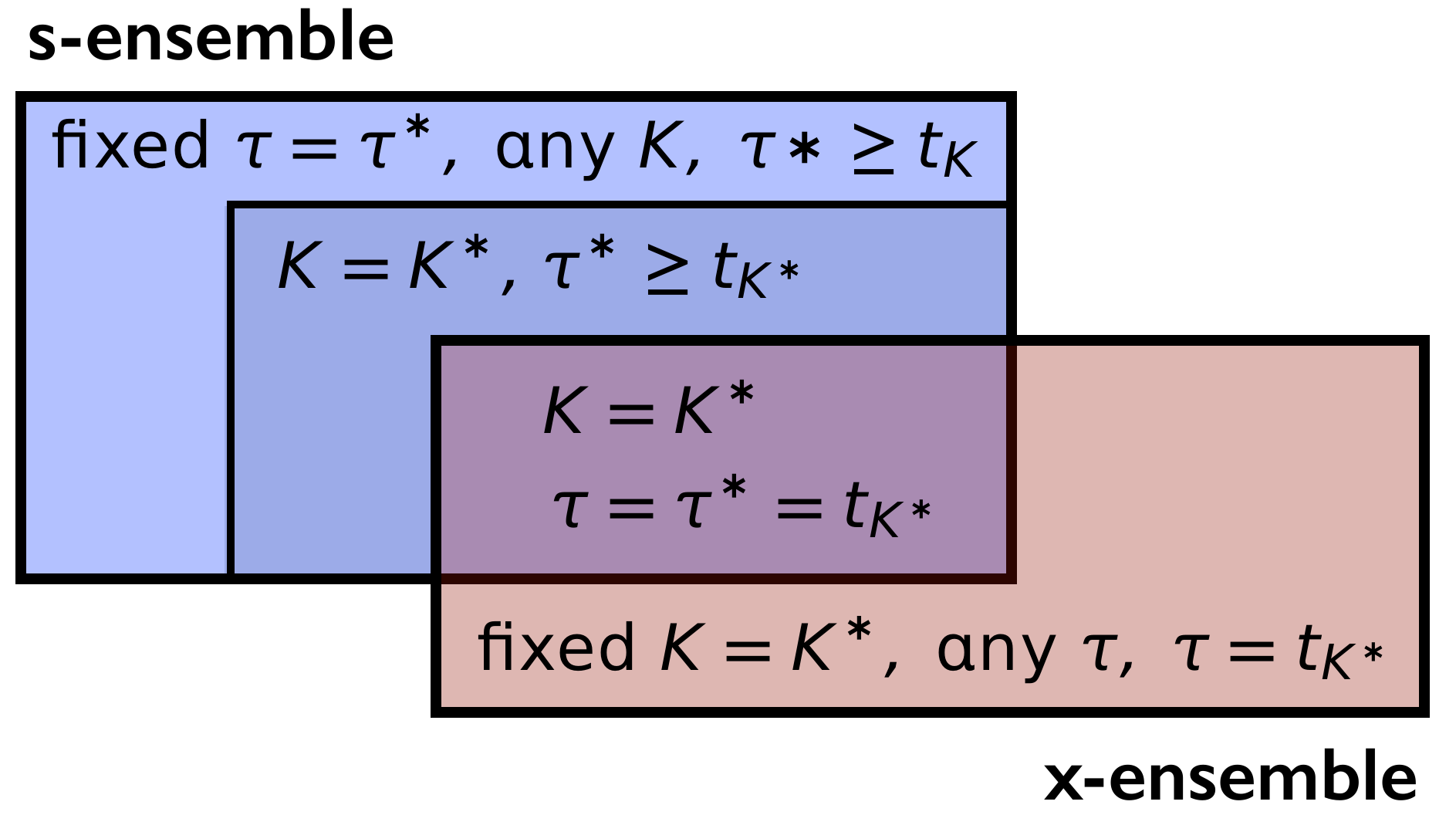}
\caption{(Color online) Sketch of concentration of measure.  The $s$-ensemble (in blue) corresponds to all trajectories with fixed overall time $\tau^*$ and any number of quantum jumps $K$, and where the time of the last jump $t_K$ is in general smaller than $\tau$.  A subset of this set is that of trajectories where the number of jumps is exactly $K^*$.  The $x$-ensemble (in red), in contrast, is that of trajectories all with fixed number of jumps $K^*$ but any time extent $\tau$ where trajectories terminate with the last jump, $\tau^*=t_{K^*}$.  
The subset of trajectories with $\tau$ exactly $\tau^*$ (in purple) also belongs to the $s$-ensemble.  The two ensembles are equivalent when concentrated to this intersection if the fields $s$ and $x$ defining the ensembles are such that $\langle K \rangle_s = K^*$ and $\langle \tau \rangle_x = \tau^*$, see Sect.\ IV.B.}
\label{fig2}
\end{figure}

\subsection{Equivalence of concentration}
Another way of formulating equivalence of ensembles is in terms of the asymptotic behaviour of the ensembles (i.e. probability distributions) themselves. In equilibrium statistical mechanics, the equivalence between canonical and microcanonical ensembles states that by appropriately scaling the extensive variable (typically energy) held fixed in the microcanonical ensemble, the canonical ensemble will concentrate around that value in the thermodynamic limit.

In order to put our result into a context, we first briefly review the corresponding classical result, proved by Touchette \cite{Chetrite2013}, using the above ``thermodynamics of trajectories" approach. (For similar results, see e.g. \cite{LPS94,Chetrite2013,CETT05}). Given an observation time $\tau$, let ${\bf X}_\tau$ denote the corresponding time truncation of any infinite trajectory ${\bf X}$, and suppose we have a family of path probabilities $\mathbb P_\tau({\bf X}_\tau)$, $\tau\geq 0$ specified by measurement process on some nonequilibrium classical system. Consider the path microcanonical ensemble $\mathbb P^{\rm mc}_{\tau;K}({\bf X}_\tau):=\delta(K-K[{\bf X}_\tau])\mathbb P_\tau({\bf X}_\tau)/P_\tau(K)$, where is $P_\tau(K)$ assumed to satisfy the LDP, i.e. $P_\tau(K):=\int d{\bf X}_\tau \delta(K[{\bf X}_\tau]-K) \mathbb P_\tau({\bf X}_\tau) \sim e^{-\tau \varphi(K/\tau)}$. Then the path canonical ensemble $\mathbb P^{\rm can}_{\tau;s}({\bf X}_\tau):=e^{-sK[{\bf X}_\tau]}\mathbb P_\tau({\bf X}_\tau)/Z_\tau(s)$ has LDP with $Z_\tau(s):=\sum_K e^{-sK} P_\tau(K)\sim e^{\tau \theta(s)}$, where \eqref{legendres} connects the potentials, we set $k(s):=-\partial \theta/\partial s$, and $K_\tau :=\tau k(s)$ for fixed $s$. Then
\begin{equation}\label{classicalEE}
\mathbb P^{\rm can}_{\tau;s}({\bf X}_\tau)/\mathbb P^{\rm mc}_{\tau;K_\tau}({\bf X}_\tau)=e^{o(\tau)}, \quad \text{for large } \tau,
\end{equation}
for those \emph{infinite} trajectories ${\bf X}$ whose time truncations ${\bf X}_\tau$ satisfy $K[{\bf X}_\tau]=K_\tau$ for each $\tau$. Here $o(\tau)$ denotes sublinear growth in $\tau$, i.e. $\lim_\tau o(\tau)/\tau =0$. In words, this result states that the canonical ensemble ``concentrates" on the microcanonical shell with $K[{\bf X}_\tau]=\tau k(s)$ \footnote{Technically, one has to take $\tau=K/k(s)$ to make $\tau k(s)$ integer.}.

In our quantum case, we can formulate an analogous statement in terms of the scalar products of the relevant quantum states. Hence, using the definitions \eqref{eq.microcanonical.K.tau} and \eqref{eq.microcanonical.tau.K}, the equivalence between the path canonical and microcanonical ensembles reads
\begin{align*}
|\langle {\rm MPS}^{\rm can}_{K;x,c}| {\rm MPS}^{\rm mc}_{K;T_K,M_K}\rangle|^2&=e^{o(K)},\\ 
|\langle {\rm cMPS}^{\rm can}_{\tau;s,c}| {\rm cMPS}^{\rm mc}_{\tau;K_\tau,M_\tau}\rangle |^2&=e^{o(\tau)},
\end{align*}

where $T_K =t(x,c) K$, $M_K=m(x,c)K$, $K_\tau=k_0(s,c)\tau$, and $M_\tau=\tilde m(s,c)\tau$. Both results can be proved easily from the definitions, using the above established LDPs. As noted before, the micro canonical states $|{\rm MPS}^{\rm mc}_{K;T_K,M_K}\rangle$ are singular vectors in the space of all trajectories with fixed $K$, but the inner product can be computed by interpreting the state as a delta distribution. Similarly, we can prove that the equivalence of the type \eqref{classicalEE} holds for the classical path probabilities $\mathbb P_K({\bf X}_K):=\langle {\bf X}|\FCS{K}|{\bf X}\rangle$ and $\mathbb P_\tau({\bf X}_\tau):=\langle {\bf X}|\cFCS{\tau}|{\bf X}\rangle$ corresponding to the fixed observables $\hat T$ and $\hat K$, respectively.

Building on the nontrivial connection \eqref{connectionxs}, we now investigate a form of asymptotic equivalence between the canonical discrete and continuous time ensembles  
$|{\rm MPS}^{\rm can}_{K;x,c}\rangle$ and respectively $|{\rm cMPS}^{\rm can}_{\tau;s,c}\rangle$. 
For given pairs $(x,c)$ and $(s,c)$ related as in \eqref{connectionxs},  we define 
$\tau_K:=Kt(x,c)=K/k(s,c)$ and $M_K:=Km(x,c)$, where $K$ is considered to be large. If $\tau := \tau_K$, then the two ensembles have the same means of extensive observables. Moreover, the quantum ensembles can be compared when restricting to trajectories for which $T({\bf X}) = \tau_K$, i.e. the evolution $V^{\tau}$ after the last jump is trivial, cf. \eqref{v.tau}. 
 

\begin{align}\label{ensemble_equiv}
& |\langle {\rm MPS}^{\rm can}_{K;x,c}|\delta((\hat T,\hat M)-(\tau_K,M_K))|{\rm cMPS}^{\rm can}_{\tau_K;s,c}\rangle|^2=\nonumber\\
&\frac{e^{-  s\cdot K - 2 c\cdot M- x\tau}P_K(\tau,M)}{Z_K(x,c) \cdot Z_{\tau_K}(s,c)}
= e^{o(K)}, \quad K \text{ large},
\end{align}
This shows that up to sub-exponential corrections, both path canonical states concentrate on the common ``microcanonical" shell where the extensive variables $\hat T$, $\hat K$, $\hat M$ have their typical values determined by the intensive quantities $t(x,c)$, $k(s,c)$, and $m(x,c)$. Note that there are only two free parameters here: we can fix either $x$ (pressure) and $c$ (in which case $s=g(x,c)$), or we fix $s$ (chemical potential) and $c$ (in which case $x=\theta(s,c)$). The relations \eqref{tkconnection} ensure that the limits $K\rightarrow\infty$ for systems of fixed $K$, and $\tau\rightarrow\infty$ for systems of fixed $\tau$ are consistent.

\subsection{Operational equivalence}

The above concentration equivalence only considers the overlap of the state on the common singular support specified by the typical values of the intensive quantities. We now proceed to formulate an operational equivalence stating that full statistics of all counting measurements up to a fixed time $\tau_0$, are the same for both states 
$|{\rm MPS}^{\rm can}_{K;x,c}\rangle$ and $|{\rm cMPS}^{\rm can}_{\tau;s,c}\rangle$, in the thermodynamic limit 
$K\rightarrow\infty$ and $\tau\rightarrow \infty$.

More precisely, we consider the scenario of long time ($\tau\rightarrow\infty$) and large number of jumps ($K\rightarrow\infty$) evolution, but only observe the output state up to a fixed time $\tau_0$ and compare the reduced output states in the two ensembles. The restriction correspond to the tensor product factorisation of the output Fock space $\mathcal F^{\rm out}$ \begin{equation}\label{split}
\mathcal F^{\rm out}=\mathcal F_{0} \otimes \mathcal F_{+},
\end{equation}
where the left (right) terms are Fock spaces of the counting fields over $(0,\tau_0]$ and $(\tau_0,\infty)$, respectively. This corresponds to the identification $|{\bf X}\rangle= |{\bf X}_{0}\rangle\otimes |{\bf X}_{+}\rangle$ of basis vectors, where any finite trajectory is split into concatenation ${\bf X}={\bf X}_{0}\vee {\bf X}_{+}$ so that ${\bf X}_{0}$ is the part with all jump times at most $\tau_0$, and ${\bf X}_+$ the rest. The trajectory integral separates as $\int d{\bf X}=\int d{\bf X}_0\int d{\bf X}_+$.

Given an arbitrary state $\varrho$ on the full output $\mathcal F^{\rm out}$, its reduced density matrix to $\mathcal F_{0}$ reads
\begin{align}\label{partialtrace}
\langle {\bf X}_{0}|{\rm tr}_{+}[\varrho]|{\bf X}_{0}'\rangle&=\int d{\bf X}_{+} \langle {\bf X}_{0} \vee {\bf X}_{+}|\varrho |{\bf X}_{0}'\vee {\bf X}_{+}\rangle,
\end{align}
where ${\rm tr}_{+}$ is the partial trace over $\mathcal F_+$. We will also need the projection onto the $N$th "layer" of the Fock space $\mathcal F_0$ which is given by 
$$
P^0_{N}=\int_{\{{\bf X}_0:K[{\bf X}_0]=N\}} d{\bf X}_0 |{\bf X}_0\rangle\langle {\bf X}_0|;
$$
these are just the eigenprojections of the number operator $\sum_{N=0}^\infty N\,P^0_N$ that defines the basic counting measurement for the time interval $[0,\tau_0]$.

We need to determine the reduced states ${\rm tr}_+[\FCS{K;x,c}]$ and ${\rm tr}_+[\cFCS{\tau;s,c}]$. Note that for $s=c=0$ we have ${\rm tr}_{+}[\cFCS{\tau;0,0}]=\cFCS{\tau_0}$, i.e. the state describing the physical evolution on $[0,\tau_0]$ remains the same for all $\tau\geq \tau_0$, as it should due to the temporal causality of the physical setup. However, \emph{the biased state ${\rm tr}_+[\cFCS{\tau;s,c}]$ does not have this property}, as we will see explicitly from the computation below. (The same phenomenon happens also in the classical case, as pointed out e.g. in \cite{JS10}.) The reduced state ${\rm tr}_+[\FCS{K;x,c}]$ is of course not expected to be remain constant with increasing $K$ even for the physical $x=0$ case, because in this case the jump times are not temporally constrained.

We will now proceed to prove the following result which holds for given $(s,c)$ and $x=\theta(s,c)$:
\begin{align}
{\rm tr}_+[\cFCS{\tau;s,c}]&\xrightarrow{\tau\rightarrow\infty} \rho^\infty_{s,c},\label{slimit}\\
{\rm tr}_+[\FCS{K;x,c}] &\xrightarrow{K\rightarrow\infty}\sum_{N=0}^{\infty}P^0_{N}\rho^\infty_{s,c}P^0_{N}\label{xlimit},
\end{align}
where the convergence is \emph{in the trace norm of the output},
\begin{equation}\label{limitstate}
\rho^\infty_{s,c}=e^{-\tau_0\theta(s,c)}{\rm tr}[\rho E_{s,c}]^{-1}S_{s,c}^{\dagger} E_{s,c} S_{s,c},
\end{equation}
and $S_{s,c}:\mathcal{F}_{(0,\tau_0)}\to \mathcal{H}$ is the operator defined by $S_{s,c}|{\bf X}_0\rangle = V^{\tau_0}_{s,c}[{\bf X}_0]\psi$, for all ${\bf X}_0$. Note that $\rho^\infty_{s,c}$ in \eqref{slimit} is a finite-rank operator that lives in the space spanned by the vectors $S_{s,c}^\dagger|m\rangle$, where $|m\rangle$ are the basis vectors of the system Hilbert space. In contrast, the limit state in \eqref{xlimit} has infinite rank because the projected states $P_N^0S_{s,c}^\dagger|m\rangle$ are all nonzero.

The interpretation of \eqref{slimit} and \eqref{xlimit} is that in the long run the reduced states of the two ensembles converge, and the limits have the same diagonal blocks with respect to the decomposition of the Fock space into 
fixed-$N$ layers. From the operational point of view, this means that in the limit, the two ensembles yield the same measurement statistics for any quantum observable on $\mathcal F_0$ that commutes with the number operator $\sum_{N} N P_N^0$ (i.e. is compatible with the counting measurement). However, while the time ensemble state may exhibit coherences between different layers, the discrete time ensemble state is block-diagonal. This inequivalence on the level of off-diagonal blocks is discussed in more detail in the next section.

We start the proof by noting that the second limit is crucially based on the equivalence of potentials \eqref{connectionxs}, and the eigenvectors \eqref{connectionFE}. We first define
\begin{align*}
V^{+,\tau}_{s,c}[{\bf X}_{+}]&=e^{-i(\tau-T[{\bf X}_+])H_{\rm eff}} J^{\tau_0}_{s,c}[{\bf X}_+],
\end{align*}
where $J^{\tau_0}_{s,c}$ is just $J_{s,c}$ with the first jump time $t_1$ replaced by $t_{1}-\tau_0$ so as to shift the starting point to $\tau_0$ where the trajectories ${\bf X}_+$ begin. (In case of an empty trajectory, we omit $J_{s,c}$ and put $T[{\bf X}]=0$ and $T[{\bf X}_+]=\tau_0$).
According to the split \eqref{split}, we have
$V^{\tau}_{s,c}[{\bf X}]=V^{+,\tau}_{s,c}[{\bf X}_+]V^{\tau_0}_{s,c}[{\bf X}_0]$, for any $\tau\geq \tau_0$. This shows explicitly how the Dyson expansion of the deformed ``evolution" factorises along this split. For the last part,
\begin{equation}\label{dyson}
e^{(\tau-\tau_0)\mathbb W_{s,c}}=\int d{\bf X}_{+}\, V^{+,\tau}_{s,c}[{\bf X}_+]^\dagger(\cdot)V^{+,\tau}_{s,c}[{\bf X}_+].
\end{equation}
The matrix elements of the total Fock state are
\begin{align*}
& \langle {\bf X}|\cFCS{\tau;s,c}|{\bf X}'\rangle =\frac{{\rm tr}[\rho V^{\tau}_{s,c}[{\bf X}']^\dagger V^{\tau}_{s,c}[{\bf X}]]}{Z_\tau(s,c)};
\end{align*}
hence using \eqref{partialtrace} and \eqref{dyson}, we get the reduced state:
\begin{align*}
& \langle {\bf X}_{0}|{\rm tr}_+[\cFCS{\tau;s,c}]|{\bf X}_{0}'\rangle\\
& =\frac{{\rm tr}[\rho V^{\tau_0}_{s,c}[{\bf X}_0']^\dagger e^{(\tau-\tau_0)\mathbb W_{s,c}}(\id)V^{\tau_0}_{s,c}[{\bf X}_0]]}{Z_\tau(s,c)},
\end{align*}
showing that
$$
{\rm tr}_+[\cFCS{\tau;s,c}]=S_{s,c}^\dagger \frac{e^{(\tau-\tau_0)(\mathbb W_{s,c}-\theta(s,c))}(\id)}{e^{\tau_0\theta(s,c)}e^{-\tau\theta(s,c)}Z_\tau(s,c)}S_{s,c}.
$$
Here $\tau_0$ is constant, and the only $\tau$-dependence is in the system operator sandwiched in the middle. Since that operator converges to $\rho_0^\infty(s,c)$ due to \eqref{LDZT}, and the system is finite-dimensional, we get \eqref{slimit}.
Note that for the case $s=c=0$, we just have $e^{(\tau-\tau_0)\mathbb W_{0,0}}(\id)=\id$, so the unobserved part of the trajectory does not contribute at all (as expected). For the biased case, we have to take the limit to get a fixed density matrix.

We now proceed to prove the second limit \eqref{xlimit}. We fix a pair ${\bf X}_0$, ${\bf X}_0'$ which determines an element of the reduced density matrix of the full state $\FCS{K;x,c}$. If ${\rm max}(K[{\bf X}_{0}],K[{\bf X}_{0}'])<K$, 
the relevant matrix elements of the total output state are of the form $\langle {\bf X}_0\vee{\bf X}_+|\FCS{K;x,c}|{\bf X}_0'\vee{\bf X}_+\rangle$. Since $\FCS{K;x,c}$ lives in the $K$th layer of the total Fock space, the element is zero unless $K[{\bf X}_0]+K[{\bf X}_+]=K[{\bf X}_0']+K[{\bf X}_+]=K$, i.e. $K[{\bf X}_{0}]=K[{\bf X}_{0}']$. 
Similarly, if ${\rm max}(K[{\bf X}_{0}],K[{\bf X}_{0}'])>K$ then the matrix element is zero since the full state has exactly $K$ excitations. If ${\rm max}(K[{\bf X}_{0}],K[{\bf X}_{0}'])=K$ then all counts happen before $\tau_0$ and the contribution to the reduced state is zero unless $K[{\bf X}_{0}]=K[{\bf X}_{0}'] =K$. Putting all the cases together we see that the reduced state is \emph{block-diagonal} with respect to the decomposition of $\mathcal{F}_0$ in Fock-layers. 

We will now fix $N= K[{\bf X}_{0}]=K[{\bf X}_{0}']$ and show that the $N$-block of the reduced state 
converges to the corresponding block $ P^0_{N}\rho^\infty_{s,c}P^0_{N}$ of the limit state state  \eqref{xlimit}. Since 
$K\to \infty$ we can assume that $N<K$ so the nonzero elements have $T[{\bf X}]=T[{\bf X}_+]$, and are given by
\begin{align*}
& \langle {\bf X}_0\vee{\bf X}_+|\FCS{K;x,c}|{\bf X}_0'\vee{\bf X}_+\rangle\\
&= \frac{{\rm tr}[\rho V^{\tau_0}_{0,c}[{\bf X}_0']^\dagger(J_{0,c}^{\tau_0}[{\bf X}_{+}]^\dagger J_{0,c}^{\tau_0}[{\bf X}_{+}])V^{\tau_0}_{0,c}[{\bf X}_0]]}{e^{x\tau_0}Z_K(x,c)e^{x(T[{\bf X}_{+}]-\tau_0)}}.
\end{align*}
Note that the part depending on ${\bf X}_0$ and ${\bf X}_0'$ is again of the same form as before, except that $s$ does not appear. We can fix that by defining $s=g(x,c)$, and noting that $V^{\tau_0}_{s,c}[{\bf X}_0]=e^{-\frac 12sN}V^{\tau_0}_{0,c}[{\bf X}_0]$. We then use \eqref{partialtrace} to integrate over ${\bf X}_+$; now the integration only goes over trajectories with $K[{\bf X}_+]=K-N$. By making the change of variables $t_i\mapsto t_i+\tau_0$ on ${\bf X}_{+}$ we can transform the integral:
\begin{align*}
&\int_{\{{\bf X}_+:K[{\bf X}_+]=K-N\}} d{\bf X}_{+} \frac{J_{0,c}^{\tau_0}[{\bf X}_{+}]^\dagger J_{0,c}^{\tau_0}[{\bf X}_{+}]}{e^{x(T[{\bf X}_{+}]-\tau_0)}}\\
&= \int_{\{{\bf X}:K[{\bf X}]=K-N\}} d{\bf X} \,J[{\bf X}]^\dagger J[{\bf X}]e^{-xT[{\bf X}]-c\cdot M[{\bf X}]}\\
&=\mathbb T_{x,c}^{K-N}(\id),
\end{align*}
and hence
$$
P^0_N{\rm tr}_+[\FCS{K;x,c}]P^0_N =
\frac{P^0_{N}\,S_{s,c}^\dagger \mathbb T_{x,c}^{K-N}(\id)S_{s,c}\,P^0_{N}}{e^{x\tau_0-sK}Z_K(x,c)e^{s(K-N)}}.
$$
Now we use the fact that $s=g(x,c)$, together with \eqref{LDZK}, to take the limit $K\rightarrow\infty$; the system operator in the middle converges, and we get
$$
P^0_N{\rm tr}_+[\FCS{K;x,c}]P^0_N \xrightarrow{K\rightarrow\infty} P^0_{N}\frac{S_{s,c}^\dagger F_{x,c}S_{s,c}}{e^{x\tau_0}{\rm tr}[\rho F_{x,c}]}P^0_{N},
$$
again in the trace norm. Since $x= \theta(s,c)$ and $F_{x,c}=\alpha_{s,c} E_{s,c}$ by \eqref{connectionFE}, the operator in the middle coincides with $\rho_0^\infty(s,c)$. To obtain \eqref{xlimit} we note that the separate convergence of the $N$-blocks implies the convergence in trace norm of the full state, which can be shown by using a quantum version of Scheffe lemma, cf. e.g. Theorem 2.20 in \cite{Simon}.

\subsection{Inequivalent aspects}

Having established that the above three levels of ``ensemble equivalence" hold, we close this section by emphasising that \emph{the two ensembles are not equivalent in all aspects.} In fact, from the statement of operational equivalence, we can already observe that there is be a difference in \emph{coherences} between different Fock layers; while the continuous time ensemble contains coherent superpositions of output states with different number of excitations, the discrete time output is a mixture of such states. A consequence of this is that the two ensembles can exhibit different properties, when the coherences are relevant. In order to see that this is indeed the case, we look at the following simple transformations of the process parameters $H$ and $L_i$:
\begin{itemize}
\item[(P1)] $H\mapsto H$, $L_i\mapsto e^{i\phi}L_i$.
\item[(P2)] $H\mapsto H+\phi\id$, $L_i\mapsto L_i$.
\end{itemize}
Here $\phi$ is an arbitrary real number. It is straightforward to verify that these correspond to the following transformations of the $x$-and $s$-ensemble states:
\begin{itemize}
\item[(P1)] $\FCS{K;x,c}$ invariant, $\cFCS{\tau;s,c}\mapsto \cFCS{\tau;s-2i\phi,c}$
\item[(P2)] $\cFCS{\tau;s,c}$ invariant, $\FCS{K;x,c}\mapsto \FCS{K;x-2i\phi,c}$.
\end{itemize}
In particular, each transformation only changes one of the ensembles. Note that the imaginary shift in the $s$-parameter changes the matrix elements of $\cFCS{\tau;s,c}$ according to
$$\langle {\bf X}|\cFCS{\tau;s,c}|{\bf X}'\rangle\mapsto e^{i\phi(K[{\bf X}]-K[{\bf X}'])}\langle {\bf X}|\cFCS{\tau;s,c}|{\bf X}'\rangle,$$ hence introducing the coherences not present in the diagonal elements; hence this inequivalence is in accordance with the above equivalence results. 
The inequivalence of the ensembles in phase transformations (P1) and (P2) has consequences for parameter estimation of Markov processes; this is however beyond the scope of the present paper, and will be analysed in a separate publication.

\section{Renewal processes}\label{sec:renewal}

As already mentioned, the discrete MPS associated to the $x$-ensemble is structurally simpler than the continuous MPS describing the $s$-ensemble. In order to illustrate the difference, we now consider a physically relevant extreme case where the former is a product state.

The measurement scheme described above is a \emph{renewal} process, if there exists a specific state $|0\rangle$ such that $L_i=|0\rangle \langle \varphi_i|$ for some vectors $\varphi_i$. This means after each measurement, the system is in state $|0\rangle$, regardless of the measurement outcome. The jump operators are $J(t_w,i)=|0\rangle\langle \varphi_i |e^{-it_wH_{\rm eff}}$, and the waiting time representation of the MPS is
$$
|{\rm MPS}_K\rangle =|0\rangle\otimes \Psi_{0,0}\otimes\cdots\otimes \Psi_{0,0}\otimes \sum_{m}\langle m|\psi\rangle \Psi_{0,m},
$$
where
$$\Psi_{0,m} =\sum_{i=1}^{N_L}\langle 0|\varphi_i\rangle \int_0^\infty dt_w \langle \varphi_i|e^{-it_wH_{\rm eff}}|m\rangle \, |t_w,i\rangle,$$
and $|m\rangle$ is a basis of the system Hilbert space containing the special vector $|0\rangle$. Hence, disregarding the first factor related to the initial state of the system, we simply have a tensor product of the single wavefunction $\Psi_{0,0}$. The transition generator is trivial, $\mathbb T(\cdot)=\id \langle 0|(\cdot)|0\rangle$, and the deformed generator is likewise not ergodic:
$\mathbb T_{x}(\cdot)=(x{\rm Id}+\mathcal R)^{-1}(D)\langle 0|(\cdot)|0\rangle$, where $D=\sum_{i}|\varphi_i\rangle\langle \varphi_i|=-2{\rm Im} H_{\rm eff}$. However, $\mathbb T_x$ clearly has only one nonzero eigenvalue $e^{g(x)}=\langle 0|(x{\rm Id}+\mathcal R)^{-1}(D)|0\rangle$. Hence, the dominant eigenvectors $\rho_x=|0\rangle\langle 0|$, and $F_x=(x{\rm Id}+\mathcal R)^{-1}(D)e^{-g(x)}$ are uniquely determined, and we see directly that the limit \eqref{LDZK} holds. Moreover, $\rho_s=(x{\rm Id}+\mathcal R_*)^{-1}(|0\rangle\langle 0|)/{\rm tr}[(x{\rm Id}+\mathcal R_*)^{-1}(|0\rangle\langle 0|)]$.

A basic example of a renewal process is the simple 3-level system with the ground state $|0\rangle$, and two exited states $|1\rangle$ and $|2\rangle$, with the transitions $|0\rangle$-$|1\rangle$ and $|0\rangle$-$|2\rangle$ driven by two resonant lasers with Rabi frequencies $\Omega_1$ and $\Omega_2$, respectively. In addition, level $|1\rangle$ decays to $|0\rangle$ with rate $\kappa$, and we detect the photons emitted in this process. Hence the Hamiltonian is $H=\sum_{m=1}^2\Omega_m(|0\rangle\langle m|+|m\rangle\langle 0|)$, and we only have one Lindblad operator $L=\sqrt{\kappa} |0\rangle\langle 1|$, i.e. $N_L=1$. One can easily check that $H_{\rm eff}$ has three distinct eigenvalues with strictly negative real parts, except for some special values of the parameters; hence the normalisation conditions \eqref{normalisation} are satisfied with the single jump operator  $J(t_w)=\sqrt{\kappa} |0\rangle\langle 1|e^{-itH_{\rm eff}}$, which maps everything to the ground state. This example was used in \cite{Garrahan2010} to demonstrate how intermittency in the jump trajectories can be related to dynamical phase transitions in the system. From the point of view of the present paper, the example is interesting because the two MPS states introduced above exhibit very different aspects of the dynamics.


\section{Conclusions}

The ``thermodynamics of quantum trajectories" formalism \cite{Garrahan2010} has been successfully employed in uncovering and analysing dynamical phase transitions in open quantum systems, through the statistical properties of an appropriately biased 
``$s$-ensemble" of quantum jumps trajectories. 
In \cite{BTG14} it was shown that the large deviations rates of the fixed-time $s$-ensemble can be deduced from those of an alternative ``$x$-ensemble" containing biased trajectories with fixed number of jumps but random time length. 

Here we have strengthened this correspondence by  lifting the ensemble equivalence to the level of the quantum states \cite{Lesanovsky2013} of the two corresponding input-output systems: a continuous-time one governed by a deformed Lindblad generator, and a discrete one whose transition operator encodes the jump-to-jump dynamics. As a first equivalence result we showed that the overlap of the continuous and respectively corresponding discrete system-output (matrix product) states decreases subexpontially in the limit of large times and counts. 
Furthermore, the restrictions of the output states to a finite time interval converge, and the limits have identical diagonal blocks with respect to decomposition of the Fock space in ``layers" of fixed number of excitations. However, a specific quantum feature which is not present in the classical set-up \cite{Chetrite2013,Chetrite2014} is that the $s$-ensemble may exhibit coherences between different layers, while the $x$-ensemble is diagonal. A consequence of this inequivalence is that certain dynamical parameters may be identified by output measurements in one ensemble but not in the other. For instance, the off-diagonal blocks of the fixed-time output contain information about the phase of the jump operator, while the fixed-jumps state does not depend on it. 

On the trajectories level, in certain situation the $x$-ensemble  \cite{BTG14} can be more amenable than the $s$-ensemble \cite{Hedges2009} to numerical simulations via paths sampling techniques \cite{Bolhuis2002}. In the same spirit, we speculate that the above quantum equivalence results can contribute towards a better understanding of dynamical phase transitions on the level of quantum states. A possible application is the extension to continuous time of the Sanov Theorem for the empirical measure of multiple successive jumps, developed in \cite{HG15}. In a different direction, the two ensembles set-up could be used to unify the existing system identification and asymptotic normality theory for discrete \cite{KG14} and continuous 
\cite{CBG} quantum Markov processes.

\

\textbf{Acknowledgement.} This work was supported by The Leverhulme Trust (Grant No.\ F/00114/BG), EPSRC (Grant No.\ EP/J009776/1) and the European Research Council under the European Union's Seventh Framework Programme (FP/2007-2013) through ERC Grant Agreement No.\ 335266 (ESCQUMA) and the EU-FET Grant No.\ 512862 (HAIRS).

\end{document}